# Interfacial temperature measurements, high-speed visualization and finite-element simulations of droplet impact and evaporation on a solid surface

Rajneesh Bhardwaj[1], Jon P. Longtin[2] and Daniel Attinger[1, *]

[1]Laboratory for Microscale Transport Phenomena,

Department of Mechanical Engineering,

Columbia University, New York, NY 10027

[2]Thermal-Laser Laboratory

Department of Mechanical Engineering,

State University of New York at Stony Brook, Stony Brook, NY 11794

*Corresponding author. Tel: +1-212-854-2841; fax: +1-212-854-3304;

E-mail address: da2203@columbia.edu

## Abstract

The objective of this work is to investigate the coupling of fluid dynamics, heat transfer and mass transfer during the impact and evaporation of droplets on a heated solid substrate. A laser-based thermoreflectance method is used to measure the temperature at the solid-liquid interface, with a time and space resolution of 100 μs and 20 μm, respectively. Isopropanol droplets with micro- and nanoliter volumes are considered. A finite-element model is used to simulate the transient fluid dynamics and heat transfer during the droplet deposition process, considering the dynamics of wetting as well as Laplace and Marangoni stresses on the liquid-gas boundary. For cases involving evaporation, the diffusion of vapor in the atmosphere is solved numerically, providing an exact boundary condition for the evaporative flux at the droplet-air interface. High-speed visualizations are performed to provide matching parameters for the wetting model used in the simulations. Numerical and experimental results are compared for the transient heat transfer and the fluid dynamics involved during the droplet deposition. Our results describe and explain temperature oscillations at the drop-substrate interface during the early stages of impact. For the first time, a full simulation of the impact and subsequent evaporation of a drop on a heated surface is performed, and excellent agreement is found with the experimental results. Our results also shed light on the influence of wetting on the heat transfer during evaporation.

*Keywords*: droplet impact, droplet evaporation, interfacial heat transfer, temperature measurement, laser temperature measurement, high-speed visualization, numerical simulation.

**Nomenclature**

| | |
|---|---|
| $c$ | drop liquid vapor concentration [kg m$^{-3}$] |
| $c_p$ | specific heat [J kg$^{-1}$ K$^{-1}$] |
| $d$ | diameter of the drop [m] |
| $D$ | diffusion coefficient [m$^2$ s$^{-1}$] |
| $h$ | heat transfer coefficient [W m$^{-2}$ K$^{-1}$] |
| $H$ | droplet height [m] |
| $k$ | thermal conductivity [W m$^{-1}$K$^{-1}$] |
| $L$ | latent heat [J kg$^{-1}$] |
| $n$ | refractive index [-] |
| $r$ | wetted radius of the droplet [m] |
| $R$ | reflectivity [-] |
| $\mathbf{v}$ | velocity vector, $\mathbf{v} = (u, v)$ |
| $V$ | Volume of the droplet [m$^3$] |
| $U$ | photodiode voltage [V] |
| $t$ | time [s] |
| $T$ | temperature [K] |

Greek letters

| | |
|---|---|
| $\alpha$ | thermal diffusivity [m$^2$ s$^{-1}$] |
| $\beta$ | gradient of surface tension with respect to the temperature [N m$^{-1}$ K$^{-1}$] |
| $\phi$ | wetting angle of the drop [-] |
| $\gamma$ | surface energy [J m$^{-2}$] |
| $\lambda$ | wavelength of laser light [m] |

$\mu$    dynamic viscosity [Pa s]

$\rho$    density of the droplet liquid [kg m$^{-3}$]

Subscript

0    ambient temperature

avg    average value

c    droplet-substrate interface

cap    spherical cap

drop    droplet

g    gas

i    initial

int    droplet-air interface

L    liquid

Ma    Marangoni effect

osc    oscillation

sub    substrate

$\infty$    far-field of the droplet

Superscript

0    equilibrium

# 1 Introduction

The impact and evaporation of a droplet on a solid substrate is relevant to a wide range of industrial processes. For instance, spray cooling [1-5] delivers heat fluxes up to 1 MWm$^{-2}$, which are orders of magnitude higher than conduction or air-cooling techniques because of the latent heat release. Other important applications are spray coating, ink-jet printing [6] and deposition of solder bumps on printed circuit boards [7, 8]. A variety of fluids are used in droplet deposition processes, e.g. water, alcohols, dielectric fluids, hydrocarbon fuels, molten metals and polymers [9, 10]. Also, processes involving the deposition and evaporation of colloidal drops can organize small structures such as proteins and DNA molecules [11, 12], micro- and nanowires [13, 14] and explosive crystalline layers [15]. The coupled and multi-scale transport phenomena occurring during the impact and evaporation of a drop on a solid substrate involve complex fluid dynamics, heat and mass transport. The fluid flow is transient within a severely deforming liquid-gas interface exhibiting evaporation as well as Laplace and Marangoni stresses, and dynamic wetting at the contact line. Heat transfer occurs by convection inside the drop and conduction in the substrate, with evaporative latent heat contribution at the evaporating liquid-gas interface and possibly an imperfect thermal contact at the drop-substrate interface. Mass transfer occurs through evaporation and vapor diffusion in the atmosphere [16]. Both numerical and experimental studies involving microdroplets are challenging. Numerical studies need to tackle the very coupled physical problem exposed above, resolving all the relevant time and size scales, in deforming domain. Experimental studies of droplet deposition need to deal with a disparate range of time scales; for example the characteristic oscillation time of a jetted drop $(\rho V/\gamma)^{0.5}$ [17], which for a 100 nL water drop corresponds to about 1 ms, while the evaporation of the same drop can take several minutes.

## 1.1 Visualization and temperature measurements

Short light pulses combined with photography and videography have been used successfully as a primary source of information for the droplet impact, spreading and evaporation [18-20]. In this method, the experiment is repeated and the delay between the drop generation and the light pulse is varied; the visualization is then achieved by piecing together the images. For instance, using flash videography with 150 ns light pulses, Attinger et al. [21] visualized the impact, spreading, oscillations and solidification of a molten solder microdroplet, a process that occurs in less than 300 μs. Another way to record the evaporation of microliter water droplets is to use a high-speed camera, as done by Mollaret et al. [22] or Wang et al. [23]. The images can then be processed to obtain evaporation parameters such as droplet volume, wetting angle, and wetted diameter. Very recently, Lim et al. [24] recorded the spreading and evaporation of picoliter droplets of water and ethylene glycol on a heated substrate using a high-speed camera .

Known limitations of such visualization methods include the *temporal resolution*, limited by the frame rate or by the conjunction of the strobe duration and jetting repeatability, and the *spatial resolution*, limited by the imaging system (lens aberrations, viewing angles, lighting, shadowing of the contact line for high wetting angles, and imaging detector pixel size/resolution). Also, the resulting images are two-dimensional, while the deformation is three-dimensional, hence some regions of the liquid-gas interface can be occluded by other portions of the liquid.

To measure temperature, pyrometers [25] and thermocouples have traditionally been used at the drop-air and the drop-substrate interface, respectively. Aziz and Chandra [26] used commercially available sheet thermocouples with a 10 μs response time to measure the temperature under impacting molten tin droplets on a stainless steel substrate. They estimated the value of the thermal contact resistance by matching the transient measured surface temperature with an analytical solution [26]. Wang and Qiu [27] measured the interface temperature during rapid contact solidification of Indalloy on a copper substrate using a micro-fabricated thermocouple with a time constant of 5 μs. Heichal et al. [28] manufactured a thin film thermocouple with a response time of 40 ns, and measured interfacial temperatures during the impact of 4 mm molten metal (aluminum alloy and bismuth) droplets with 1-3 m s$^{-1}$ impact velocity [29]. They obtained the value of thermal contact resistance during the early stages of the spreading by matching measured surface temperature with an analytical solution of the one-dimensional transient heat conduction equation [29].

The spatial resolution of these thermocouples is relatively large compared to their temporal resolution, as thin-film thermocouples usually have spatial resolutions on the order of several millimeters. Also, thermocouple measurements are intrusive and can disturb the fluid flow and heat transfer through physical contact and by altering the surface wetting properties. While most thermocouple studies measured the temperature at the center of impact, an attempt was made to measure the spatial variation of temperature along the interface using an array of 96 feedback-controlled heaters with spatial and temporal resolutions of respectively 8 μm and 1 μs, Kim et al. [30, 31]. They reported the measurement of local wall heat flux during the impact of a FC-72 droplet. Feedback loops were used to vary the voltage across each heater in the array to keep its temperature (and thus resistance) constant. In a similar study [32], Xue and Qiu manufactured an array of 64 MEMS temperature sensors. They probed the temperature between a droplet and a glass substrate, with a spatial and time resolution of 50 μm and 100 ns, respectively. Very recently, Paik et al. [33] measured the temperature at the drop-substrate interface for evaporating water droplets on glass substrate using an array of 32 microheaters, at a frequency of 100 Hz.

Despite these many efforts, a need remains for a *non-intrusive* temperature measurement at the drop-substrate interface, with both high time and space resolution. Laser-based techniques have the advantages of being non-intrusive and having high spatial and temporal resolutions. In 2003, Chen et al. [34] presented a non-contact, laser-based thermoreflectance technique to measure the time-dependent solid-liquid interface temperature. This method is based on measuring the change of reflectivity due to the temperature-dependent index of refraction at the substrate-liquid interface (see Figure 1). The thickness of the interrogated liquid region is on the order of half a wavelength, corresponding to a very close estimate of the true interface temperature. They measured the temperature during the impact of heated drops on a colder substrate, with time and spatial resolutions of 8.8 ms and 180 μm [34], respectively. In the present paper, a laser measurement system based on the same thermoreflectance principle is implemented with improved time and space resolution of respectively 100 μs and 20 μm. A high-speed camera (Redlake HG-100K) is simultaneously used for recording the impact, at frame rates up to 3000 per second. It is worth mentioning that the laser-based technique used in this work can also be used for multipoint temperature mapping along the droplet-substrate interface. To do this, experiments can be repeated with either the impact location or the sensing location moved slightly between each experiment. However, we did not perform multipoint temperature mapping in this work.

## 1.2 Theory and numerical studies

Due to the severe coupling of transport phenomena, early numerical models of droplet impact on a solid surface were simplified for the sake of numerical tractability, neglecting, e.g. viscosity or surface tension [35-37]. In the late 1990s, more accurate results were obtained with the volume-of-fluid (VOF) method [38-42]. Pasandideh-Fard et al. [41] obtained good agreement between VOF simulations and photographs of tin droplets impacting on a steel plate. Measured values of advancing wetting angle were used as a boundary condition for the numerical model. Other methods involving interface reconstruction on a fixed grid are the *level-set method* [43] and the use of markers in the front-tracking approach in Tryggvason's group [44]. An alternative to methods involving interface reconstruction is the use of a Lagrangian approach, which was developed by Fukai et al. [17] to investigate the droplet impact process. In this approach, the mesh moves with the fluid, allowing precise tracking of the deforming free surface. The modeling in [17] was made with the Finite Element method and solved the full Navier-Stokes equations. The simulations predicted realistic features such as the formation of a propagating ring structure (due to mass accumulation) at the periphery of droplet, as well as recoiling and subsequent oscillation of the droplet. The modeling in [17] was extended by Zhao et al. [45] for heat transfer to simulate the cooling of liquid metal (tin and aluminum) and water droplets, and found that heat transfer between an impinging microdroplet and a substrate at a different initial temperature occurs simultaneously with spreading. Subsequently, Waldvogel and Poulikakos [46] extended the model to account for solidification and modeled the impact of solder droplets on composite two-layer substrates. Their results documented the effects of impact velocity, initial droplet diameter, thermal contact resistance and substrate thermophysical properties. Pasandideh-Fard et al. [39] and Xiong et al. [47] performed a numerical study on the sensitivity to interfacial heat transfer coefficient on the final diameter, overall shape and height of a solidified solder droplet. Their model predicted variations in solder bump height up to 20% due to variations of interfacial heat transfer coefficient. Attinger and Poulikakos [48] compared the numerical and experimental amplitude and frequencies of oscillations of a solidifying solder drop and were able to estimate the value of the interfacial heat transfer coefficient between the drop and the substrate for a specific case. Recently, Xue et al. [28] modeled thermal contact resistance as a function of substrate roughness and thermal conductivity. This modified numerical model [28] was able to accurately predict the impact dynamics of 4 mm diameter aluminum alloy droplets landing on a tool steel plate with 3 m s$^{-1}$ velocity. Regarding droplet evaporation, several numerical and analytical studies have been reported. Using the lubrication approximation, Deegan et al. [16, 49, 50] and Hu and Larson [51, 52] showed analytically that pinning together with maximum evaporation at the wetting line creates a steady flow from the droplet interior to the wetting line. Also, a detailed numerical model for the evaporation of pure microliter water drops was presented by Ruiz and Black [53]. They solved the Navier-Stokes and energy equations with consideration of thermocapillary stresses, for a pinned contact line. Studies by Mollaret et al. [22], Girard et al. [54] and Widjaja et al. [55] followed a similar approach assuming a pinned wetting line but solved the vapor diffusion equation to obtain an accurate evaporative flux. Due to the complex, coupled physics involved during drop evaporation, most theoretical and numerical models reported so far are based on assumptions such as fluid flow with negligible inertia [49, 51, 56, 57], small wetting angle [16, 49], spherical cap shape of the free surface [16, 22, 49, 51, 56-59], pinned wetting line throughout the evaporation [22, 49, 51, 53, 54, 56, 57, 60-62], negligible heat transfer between the drop and the substrate [16, 49-51, 61, 62] and negligible Marangoni convection [55, 61, 62]. In this study, we compare the experiments with a model based on the Lagrangian approach in [46], solving the Navier-Stokes equations and convection and conduction heat transfer equations. Our model allows for free surface deformation and receding at the wetting line, as observed in experiments [22, 58, 63-65]. The modeling is described in more details in section 3.

In this paper, we study the fluid dynamics and heat transfer during the impact of microliter drops on a flat, heated solid substrate, and during the deposition and evaporation of nanoliter drops on a flat heated solid substrate. To do so we combine the following techniques:

- thermoreflectance measurement of the temperature at the drop-substrate interface
- high-speed visualizations
- numerical simulations

Isopropanol is chosen as the working fluid because of its large change in refractive index with temperature. Section 2 of the article describes the experimental method, section 3 describes the numerical model, and section 4 describes experimental and numerical results for the impact and the evaporation of respectively microliter and nanoliter isopropanol drops.

## 2 Experimental setup

The configuration of the experimental setup is shown in Figure 1. Microliter and nanoliter isopropanol droplets are generated using, respectively, a standard syringe with 26.5 gage needle and a fast solenoid valve with a ceramic nozzle of 420 μm diameter connected to a pressurized reservoir. A 90° fused silica prism with 35 mm edge length serves as the heated impact surface (refractive index ~ 1.457 at 25$^o$C [66]). The prism is sandwiched between a thermoplastic PEEK GF-30 plate (max operating temperature 350$^o$C) and another PEEK plate held by three screws with embedded spring-loaded plungers to compensate for the thermal dilatation. The temperature of the prism is controlled from ambient to 130°C using a temperature controller and a K-type thermocouple for temperature feedback. The prism is held between two aluminum plates in contact with its two triangular faces, in a spring-loaded PEEK holder able to accommodate thermal dilatation and temperatures up to 150$^o$C. Two cartridge heaters, in each aluminum plate, help maintain a uniform temperature at the center of the impact surface. The substrate temperature is measured with a pyrometer (0S-611-A, Omega Inc, resolution 1$^o$C). The prism provides high transmittance in the visible spectrum and a smooth, flat and horizontal surface for the solid-liquid interface. Before each drop impact, the substrate is cleaned with methanol and allowed to dry for one minute.

The temperature at the center of the drop-substrate interface is measured with the laser thermoreflectance technique (Figure 1b) extensively described in [34] and summarized here. In this method, the temperature change at the interface can be expressed as a function of the reflectivity ($R = I_3/I_0$), the voltage from the photodiode/amplifier $U$, and the change in reflectivity with temperature as follows:

$$\Delta T = \left[ \frac{\partial R}{\partial n_g} \frac{dn_g}{dT} + \frac{\partial R}{\partial n_L} \frac{dn_L}{dT} \right]^{-1} R_0 \frac{\Delta U}{U_0} \quad (1)$$

In the above equation, $U_0$ and $R_0$ are the voltage and reflectivity measured at ambient temperature $T_0$, respectively. The sensitivity of the measurement is defined as $\Delta R/R_0$ for $\Delta T = 1$ K. In order to provide the smallest spot size and therefore highest spatial resolution, the beam is focused using a Galilean telescope consisting of one diverging and two converging lenses, shown in Figure 1. The beam strikes one side of the fused silica prism at an angle $\alpha = 8°$ from normal incidence. The prism material, geometry and the beam angle were chosen to provide a good sensitivity to fluid temperature variations, while having a relatively small angle $\gamma$ at the glass-air interface so that the spot diameter can be measured using a knife-edge technique [67]. For comparison the prism material and geometry used in [34] did not allow the measurement of the spot size because of total reflection of the beam at the glass-air top surface of the prism. The spot diameter on the impinging surface was measured to be 21 μm. For water drops, values of $\gamma$ and the sensitivity were calculated respectively as 42° and $1.68 \times 10^{-3}$. For isopropanol, $\gamma$ is 59.5° and the sensitivity of $10.5 \times 10^{-3}$ was almost one order of magnitude larger than the one of water, owing to the larger temperature dependence of the isopropanol index of refraction (see Table 1). S-polarized light was used for all experiments because of a larger transmission and more linear sensitivity than p-polarization. The temperature was kept below the boiling point of isopropanol, 82.3°C.

A 1mW 632.8 nm HeNe laser (05-STP 901, Melles Griot) was used as the light. This wavelength was chosen because of the availability of a stabilized laser in our laboratory. The power stability in intensity-stabilized mode is ±0.1% rms, and measured stability at the diode after the heated prism was ±1% rms. After reflection on the liquid-glass interface, the beam is focused through a plano-convex lens onto a Melles Griot 13DSI003 planar-diffused silicon photodiode of 3.6 mm diameter. The photodiode current is converted with a Melles Griot 13AMP005 transimpedance amplifier to a voltage that is linearly proportional to the photodiode current. Since only the reflected beam from the liquid–solid interface is of interest, a spatial filter with a 40 μm pinhole is used to stray light, e.g., reflected beam from the liquid-air interface, from striking the photodiode. A PC with a National Instruments PCI-5911 data acquisition system is used to acquire the amplifier output voltage at a sampling rate of 10 kS s$^{-1}$. The calibration was done by placing a 6 mm high elastomer ring on top of the prism, filled to create a liquid pool. The calibration curve for converting the photodiode voltage to temperature is shown in Figure 2. The horizontal axis in Figure 2 is the temperature measured with a calibrated K-type thermocouple in the liquid half a millimeter from the pool bottom. The vertical axis is the reflectivity measured relatively to the reflectivity at ambient temperature. Since the temperature sensitivity of the index of variation of fused silica is thirty times smaller than that of isopropanol, it follows from equation 1 that the measured interfacial temperature is the temperature of the liquid at the interface.

To minimize noise, the experiment is assembled on a vibration-isolated optical bench. The measurement uncertainty is mainly due to the uncertainty in the laser power (±1% rms), resulting in an uncertainty of ±1$^{\circ}$C for the isopropanol drops. Correspondingly, the measurement uncertainty for water drops was about ±6$^{\circ}$C due to the roughly six times lower sensitivity, hence results for water are not reported here.

High-speed visualizations of the droplet impact are performed using a Redlake MotionXtra HG-100K digital camera and zoom objective. The chosen magnification corresponds to 6 μm per pixel, resulting in a spatial uncertainty of ±6 μm. Both the exposure time and image size of the camera are adjustable, allowing for bandwidth maximization. Typically, a frame rate and image size of respectively 3000 frames per second and 840×480 pixels were used, for the impact case, with lower frame rate for the subsequent evaporation. Experimental images obtained are analyzed using NIH image software [68]. Based on the measured height, $H$ and the wetted radius, defined as the distance between wetting line and center of the symmetry of the drop, $r$, in Figure 4, the drop volume $V$ and the wetting angle $\phi$ are calculated during the evaporation using the following equations for a spherical cap: $V = (1/6)\pi H(3r^2 + H^2)$ and $\phi = 2\tan^{-1}(H/r)$.

## 3  Numerical model

The mathematical model used is based on an in-house finite-element code that solves the full Navier-Stokes equations for droplet impact [69]. The model also solves convection and conduction heat transfer in the droplet and substrate. This model also accounts for evaporation and the associated heat transfer at the drop free surface [69], and the variation of viscosity and surface tension with temperature. A tunable interfacial heat transfer coefficient is used to model heat transfer at the drop-substrate interface. The code was developed by Poulikakos and co-workers [35, 36, 46, 48, 70], and by Attinger and co-workers [69, 71, 72]. This model has been extensively validated for studies involving impact and heat transfer of molten metal drops [48, 70], of water drops [72] and colloidal drop evaporation [69]. The flow inside the droplet is assumed to be laminar and axisymmetric. All equations are expressed in a Lagrangian framework, which provides accurate modeling of free surface deformations and the associated Laplace [17] and Marangoni stresses [69]. The kinetic wetting model of Blake and de Coninck [73] is also implemented as described in [72], kinetics parameter and wetting angle determined as to match the experiments. The thermophysical properties of isopropanol and fused silica used in the simulations in this paper are given in Table 1.

Table 1: Thermophysical properties used in the simulations at 25°C [74, 75]

| Substance | $\rho$ | $k$ | $c_p$ | $\mu$ | $\gamma$ | $\beta$ | $L$ | $n^*$ | $\frac{\partial n}{\partial T}$ |
|---|---|---|---|---|---|---|---|---|---|
| Isopropanol | 780 | 0.13 | 2560 | $2.04\times10^{-3}$ | $2.09\times10^{-2}$ | $-7.89\times10^{-5}$ | $768\times10^3$ | $1.3742^e$ | $-3.9\times10^{-4\,a}$ |
| Water | 997 | 0.607 | 4180 | $9.0\times10^{-4}$ | $7.2\times10^{-2}$ | $-1.68\times10^{-4}$ | $2445\times10^3$ | $1.331^b$ | $-0.8\times10^{-4\,b}$ $-1.04\times10^{-4\,c}$ |
| Fused silica | 2200 | 1.38 | 740 | - | - | - | - | $1.457^d$ | $-1.28\times10^{-5\,d}$ |

*at $\lambda$ = 632.8 nm

$^a$Ref [76], $^b$Ref [77], $^c$Ref [78], $^d$Ref [66], $^e$Ref [74]

## 4 Results and Discussions

In this section, we study the impact and evaporation of isopropanol drops on a fused silica substrate. We generated drops of two volumes: microliter drops from a syringe and nanoliter drops from a high-speed solenoid valve. The microliter drops were used to study the fluid dynamics and heat transfer during the impact stage. The nanoliter drops were used to study the heat transfer and fluid dynamics during the impact and subsequent evaporation process.

**4.1  Fluid dynamics and heat transfer during the impact of microliter drops on a heated substrate**

The impact of 3 μL isopropanol drops, a volume corresponding to 1.8 mm diameter, on heated fused silica substrate was investigated. The initial temperature of the ejected drop was at ambient temperature 23°C. The impact velocity was set to 0.37 m s$^{-1}$ by the height of the dispensing syringe. Using the parameters at the moment of impact, Reynolds and Weber numbers are respectively 255 and 9.2. Impacts on substrates at initial temperatures of 25, 56 and 68°C were investigated. The left column of Figure 3 shows a sequence of high-speed visualization images during impact on a substrate initially at 68°C. We see that the impact phase occurs in about 5 ms, a time that approximately matches the ratio of the drop diameter over the impact velocity [38]. During the impact phase, the free surface of the drop experiences strong deformations, which are signature of the transfer of inertial energy into surface energy. Between 5 and 10 ms, the drop assumes a doughnut-like shape, with a visible depression at its center. This deformation can also be attributed to the competition between inertial and surface forces. Figure 3 and two measurements in Figure 4 show that from 10 to 40 ms, the wetted radius and droplet height oscillate, until the drop attains a wetted radius of 1 to 1.1 times the initial droplet diameter, in qualitative agreement with analytical models as in [38]. The free surface oscillations have a measured period of 15 ms. They are likely due to the competition between inertia and surface tension, the period of which can be estimated [79] as $t_{osc} = \sqrt{\rho d_i^3 / \gamma}$ = 14.75 ms, where $\rho$, $d_i$ and $\gamma$ are density, initial diameter and surface tension, respectively.

Measurements of the temperature at the glass-liquid interface are shown in Figure 5 for three drops impacting under the same conditions as Figure 3. The interfacial temperature is initially measured to be 58°C, and within 6-8 ms decreases by about 2°C. At this point, which corresponds to the end of the rapid phase of the impact (see Figure 3), all three measurements show a large temperature spike. Very likely this is an artifact due to the depression at the center of symmetry of the drop bringing the liquid-air interface very close (a few micrometers) to the prism, as inferred from Figure 3 and the numerical simulations described hereafter. In this configuration, some of the laser beam transmitted into the liquid is reflected at the liquid-air interface and strikes the photodiode, inducing an artificial increase of reflectivity. From about 10 ms to 40 ms, the temperature increases to ~59°C and then decreases again to reach 56°C at about 100 ms.

The influence of the initial substrate temperature $T_{sub}$ is studied in Figure 6 and Figure 7, which compares measurements of wetted radius, maximum height and interfacial temperature for initial substrate temperatures of 25, 56 and 68°C, all other parameters being kept constant. In Figure 6, it appears that the largest oscillations of droplet height and wetted radius are obtained for higher substrate temperatures, probably because of the corresponding lower liquid viscosity. The impact dynamics however are similar at each temperature, with final wetted radii that are comparable, being in the range of one to 1.1 times the droplet initial diameter. The wetting angles at 100 ms for the three cases ($T_{sub}$ = 68, 56 and 25°C) are measured as 30°, 28° and 26°. In Figure 7, all temperature measurements (three measurements for each temperature) exhibit spikes at a time of 6-10 ms, which confirms that during impact the top of the drop comes close to the substrate and perturbs the reflectivity measurement. The flat curve for the ambient temperature case indicates that the measurement method is not sensitive to other fluid dynamic events. Interestingly, the initial interfacial temperature $T_c$ measured at the interface can be estimated by assuming that two semi-infinite bodies are suddenly put in perfect contact, as follows:

$$T_c - T_{drop,i} = \frac{b_{sub}}{b_{sub} + b_{drop}} (T_{sub,i} - T_{drop,i}) \qquad (2)$$

where $b_{sub}$ and $b_{drop}$ are the respective effusivities $(k\rho c_p)^{1/2}$ of the substrate and the drop, and $T_{sub,i}$ and $T_{drop,i}$ are respectively the initial substrate and drop temperature. This analysis is valid at the instant of impact only, when the thermal penetration depth is negligible with respect to the size of both the droplet and substrate [34, 80, 81]. Using the above equation, estimates of the initial interface temperature are respectively 56.4, 47.5 and 24.5°C for respective initial substrate temperatures of 68, 56 and 25°C. These temperatures compare well with the initial measured values of respectively 58, 47.5 and 25.5°C. Another interesting result shown in Figure 7 is that every temperature measurement for the cases $T_{sub}$ = 56°C and 68°C shows a non-monotonic evolution of the temperature at the center of the drop-substrate interface, with an initial decrease for the first 10 ms, and then an increase until about 40 ms. This oscillating behavior will be explained in the next paragraph, which compares the simulation results with the measurements.

Indeed, numerical simulations can explain several of the trends observed experimentally. Simulated temperatures and instantaneous streamlines of the flow field are shown in Figure 3, where care is taken to have the same scaling for the experimental and numerical images, with the substrate level shown on the right of the simulation results. We used a kinetic parameter $K_w = 1 \times 10^7$ Pa and an equilibrium wetting angle $\phi^0 =$ 30° in the wetting model to match the experiments. The agreement between the measured and simulated droplet surface shapes is very good, except for a significant difference at 7 ms, when the simulated drop height on the axis of symmetry seems lower than the measured one. The reason is that the experiment shows an image taken from the side, while the simulation is a radial cut. Simulations show that the outward flow reverses between 7 and 10 ms, confirming that the deposition is a competition between inertia and surface tension forces. From 10 to 15 ms, the main flow direction is from the edge towards the drop center, inducing an increase of the drop height and a donut-shaped vortex centered at the bottom of the drop. For times larger than 20 ms, the flow reverses again as the height of the droplet decreases, in an oscillation process driven by energy exchanges between the surface energy and the kinetic energy. In Figure 4, the agreement between the measured and simulated evolution of the wetted radius and the droplet maximum height is very good, with similar impact dynamics, oscillation frequencies and final radii and heights. In Figure 3, the visualization at 20 ms show a very thin film protruding from the apparent wetting line, a phenomenon not captured by the numerical model. The simulations also explain several heat transfer trends. First, the temperature oscillations measured under the drop in Figure 5 and Figure 7 can be explained by the convection associated with the impact and oscillations of the drop. During the first 8 ms the drop impacts the substrate (see simulations in Figure 3), providing a steady supply of colder fluid that monotonically decreases the substrate temperature, as shown by the direction of the streamlines and temperature contours in Figure 3. During the later phase (10 to 40 ms), the temperature field and streamlines visible in Figure 3 show the fluid previously heated during the radial impact being brought back to the droplet center, and subsequently slowing down, so that heating from the substrate causes an increase in the fluid temperature. The switch from a rapid impact with convection to a flow reversal and deceleration explains the temperature oscillations observed between the impact and 40ms. Further simulations in Figure 5 also indicate that the oscillations of interfacial temperature seen during the impact phase (0 – 30 ms) only occur for a high interfacial heat transfer coefficient ($h_c$ = 1.5×10$^5$ W m$^{-2}$ K$^{-1}$ and above). For instance, the simulation with the lowest value of $h_c$ = 1.5×10$^4$ W m$^{-2}$ K$^{-1}$ shows an almost monotonic temperature increase during the impact phase. These results indicate that matching between experimental and numerical results can help determine interfacial heat transfer coefficients, at least in an order of magnitude sense. Interfacial temperatures from simulations are also plotted in Figure 7 for different values of interfacial heat transfer coefficients. The oscillations of the numerical and experimental curves are in very good agreement, with values comparable within ±3°C, provided a heat transfer coefficient $h_c$ = 1.5×10$^5$ W m$^{-2}$ K$^{-1}$ or higher is considered. Another significant contribution of the numerical simulations is to show the initiation of Marangoni convection. Figure 3 shows that by 30 ms, a donut-shape vortex has started to develop, characteristic of Marangoni convection. The counterclockwise direction of the loop can be explained by the fact that the heated drop has its wetting line warmer than its top: Marangoni stresses acting from low to the high surface tension regions pull liquid from the wetting line to the droplet top, while the strongest evaporation at the wetting line creates an internal flow field from the drop center to the wetting line. The speed of the Marangoni convection is indicated in Figure 3, and it fully develops at a time of 50 ms.

**4.2　Evaporation of nanoliter isopropanol drops on a heated fused silica substrate**

In this section, we study the fluid dynamics and heat transfer during the evaporation of nanoliter isopropanol drops on a heated fused silica substrate. We generated drops from a pressurized solenoid valve with volumes ranging from 60 to 90 nL, corresponding to initial diameters of 486 and 586 μm, and impact velocities in the range of 0.3–0.6 m s$^{-1}$, all respectively. These conditions correspond to Reynolds and Weber numbers in the respective range of 56–127 and 1.6–7.4. The relatively large range of initial conditions was due to repeatability issues with our drop generation process, but for each result presented hereafter the initial volume and velocity was precisely measured from the high-speed visualizations. The substrate initial temperature was varied from 25°C to a maximum 63°C. At these temperatures phase change occurs by evaporation at the liquid-air interface. The initial temperature of the ejected drop is at ambient temperature 23.5°C.

Figure 8 shows the typical evolution of the wetted radius and the volume during the impact of two drops on a substrate initially at 41°C, plotted with a logarithmic time scale. Results are non-dimensionalized with respect to the initial diameter and volume. The wetted radius quickly increases to a value of about 1.2 times the initial diameter for the first hundred microseconds, driven by the inertia of the impact. Then the radius of the spherical cap increases at a slower pace, driven mainly by the wetting at the contact line. At 0.5 to 1 s, receding starts, driven by the drop evaporation. The total evaporation time is about 5 s. Figure 9 shows interfacial temperature measured for the same cases, with initial substrate temperature $T_{sub}$ of 25, 41, 49, and 57°C. Also, the final evaporation time measured from the movies is indicated as a vertical line. For each case except the case at 25°C, the measured temperature first increases exponentially for ~ 0.5 s to reach a maximum. Then the temperature decreases linearly by about one degree until a very sudden increase occurs. This sudden temperature jump is due to the drop becoming a thin film at the end of the evaporation, inducing parasitic reflection of the laser beam at the liquid-air interface. For each case the temperature in the second stage is within one or two degree of the initial substrate temperature. A typical temperature measurement for the case when the substrate is at 25°C is also reported (Figure 9). It shows unusually large voltage fluctuations after 0.25 s of the drop impact. The reason is that the drop spreads like a film after 0.25 s. This thin film alters the reflectivity of the interface, resulting in strong fluctuations on the output voltage. If the fluid dynamics in Figure 8 are compared with the temperature measurements in Figure 9, it appears that the transition between the two stages of the temperature evolution (exponential increase and linear decrease) occurs simultaneously with the incipience of receding, measured in Figure 8. The reasons are that in the first stage, from $t = 0$ to 0.5 s, the heat transfer time scale is much slower than the impact time, so that the drop quickly assumes a spherical cap shape with a temperature increasing exponentially due to heat diffusion from the substrate. Indeed, the time scale for diffusion can be estimated as $t_{\text{diffusion}} = H^2/\alpha = 0.2$ s, where $H$ is the height of the spherical cap ($H \sim 100$ μm), much longer than the impact time which scales as $d_i/v_i \sim 1$ ms. During the second stage ($t = 0.6$ to 4.2 s), the evaporating drop shrinking and receding at the wetting line causes the temperature to decrease slightly; this can be explained by considering an analytical model of Popov [57, 82], assuming a spherical cap. This model establishes that the rate of change of the droplet volume $V$ equals the product of the wetted radius $r_{cap}$, a geometric function $f(\phi)$ of the wetting angle, the diffusion coefficient and the vapor concentration difference between the liquid-air interface and the environment.

$$\rho \frac{dV}{dt} = -\pi r_{cap} D(c_{\text{int}} - c_{\infty}) f(\phi) \tag{3}$$

where $\rho$ is the density of the droplet liquid, $D$ is the vapor-phase diffusivity of the droplet liquid in air [m$^2$ s$^{-1}$], and $c$ is the drop liquid vapor concentration [kg m$^{-3}$]. From an energy balance, the rate of change of the droplet volume $V$ can also be expressed as the product of an overall heat transfer coefficient $h$, the wetted area $\pi r_{cap}^2$ and the temperature difference between the droplet and the substrate.

$$\rho L \frac{dV}{dt} = -h\pi r_{cap}^2 (T_{sub} - T_{drop,avg}) \quad (4)$$

where $L$ is the latent heat of the droplet liquid. Comparing the two above equations, one obtains that

$$(T_{sub} - T_{drop,avg}) = \frac{LD(c_{int} - c_\infty)}{h} \frac{f(\phi)}{r_{cap}} \quad (5)$$

with the factor $D(c_{int} - c_\infty)$ depending only weakly on temperature. Since receding occurs at constant wetting angle, $f(\phi)$ is correspondingly constant, and the decrease of $r_{cap}$ during receding can explain the observed temperature decrease.

Several of the trends observed experimentally can be explained with the help of numerical simulations for the same conditions. For all evaporation simulations a perfect thermal contact is assumed at the interface. We used a kinetic parameter $K_w = 5\times10^6$ Pa and an equilibrium wetting angle $\phi^0 = 30°$ in the wetting model to match the experiments. The case simulated in Figure 8 is the impact of a 74 nL isopropanol drop on a fused silica substrate heated at 41°C. The initial diameter, velocity before the impact and receding angles are 0.4 m s$^{-1}$, 521 μm and 10°, as measured from the visualization frames. This case corresponds to Reynolds and Weber number of respectively 80 and 3.1. To the best of our knowledge it is the first time a simulation is made that describes the entire evaporation phenomenon, from the drop impact to its full evaporation. The temperature and flow fields are shown in Figure 10, with the simulated radius and volume plotted in Figure 8. The first phase of the spreading involves strong deformations of the free surface, and at 2.7 ms the drop has already assumed the shape of a spherical cap. Then, the drop spreads and evaporates simultaneously until approximately 3 second, when receding starts (Figure 8). Figure 10 shows that it takes approximately 40 ms to establish a Marangoni convection loop, due to the combined effect of the radially outward flow driven by the evaporation and the free-surface Marangoni stresses. The maximum speed of the Marangoni convection is shown in Figure 10 at $t = 1$ s. An analytical estimate of the Marangoni convection speed is given by Hu and Larson [60]:

$$\|\mathbf{v}_{Ma}\| \sim \frac{1}{32} \frac{\beta \phi_i^2 \Delta T}{\mu} \qquad 6$$

where $\phi$ is the wetting angle of the drop, $\mu$ is dynamic viscosity [Pa s], $\beta$ is the gradient of surface tension with respect to the temperature, and $\Delta T$ is the temperature difference between the edge and the top of the droplet. For the case shown in Figure 10, the Marangoni speed obtained analytically is on the same order as in our simulations ( $\|\mathbf{v}_{Ma,analytical}\| \sim 1.3 \times 10^{-4}$ m s$^{-1}$ and $\|\mathbf{v}_{Ma,numerical}\| \sim 7.3 \times 10^{-4}$ m s$^{-1}$). Interestingly, the simulation shown in Figure 10 is to the best of our knowledge the first simulation that describes both the impact and the evaporation of a drop. During most of the drop evaporation, the temperature contours shown in Figure 10 are horizontal ($t > 1$ s), showing the dominance of conduction heat transfer over convection. The numerical evolution of the temperature at the drop-substrate interface is shown in Figure 9. The numerical temperature shows a similar trend as the experimental temperature, first increasing exponentially until 0.6 ms and then showing a slight linear decrease. The numerical temperatures are about 2-3°C lower than the experimental ones, within the range of uncertainty of the experimental method. Numerical simulations without Marangoni convection are also reported in Figure 9. They show an evaporation time significantly larger than the ones with Marangoni convection, confirming that Marangoni convection accelerates the evaporation.

# 5 Conclusions

The impact of microliter isopropanol drops and the evaporation of nanoliter isopropanol drops on heated fused silica substrate were experimentally and numerically investigated. A laser-based method was used to measure the temperature at the solid-liquid interface. This method is based on a thermoreflectance technique and provides unprecedented temporal and spatial resolutions of 100 μs and 20 μm. High-speed visualizations of the droplet impact and evaporation were also performed. An in-house numerical code was used to simulate the fluid flow and temperature fields during the impact, wetting and evaporation phase. The modeling solves the full Navier-Stokes and heat and mass transfer equations. Comparison of measurements and simulations explain non-intuitive trends, such as oscillations of interfacial temperatures during the impact of microliter drop. Also, the modeling shows the influence of imperfect thermal contact and substrate initial temperatures on the drop impact. For the evaporation of nanoliter drops, measurements show that the interfacial temperature first increases exponentially and then decreases linearly. These two stages are shown to be influenced by the fluid dynamics of the drop, in which the initial spreading is driven by inertial and wetting forces while the subsequent receding is due to evaporation. The influence of the Marangoni effect and the substrate temperature were quantified. In general, comparisons of the numerical results with the experimental data are very good and provide insight into the complex coupling of fluid dynamics and heat transfer.

# 6 Acknowledgements

The authors gratefully acknowledge financial support for this work from the Chemical Transport Systems Division of the US National Science Foundation through grant 0622849. The authors also thank Joseph Coleman and Xiaolin Wang for contributing to the design and setup of the laser thermoreflectance measurement.

[58] R. G. Picknett and R. Bexon, "The Evaporation of Sessile or Pendant Drops in Still Air," *Journal of Colloid and Interface Science*, vol. 61, pp. 336-350, 1977.
[59] J. J. Dyreby, G. F. Nellis, and K. T. Turner, "Simulating fluid flow in lithographically directed, evaporation driven self assembly systems," *Microelectronic Engineering*, vol. 84, pp. 1519-1522, 2007.
[60] H. Hu and R. G. Larson, "Analysis of the Effects of Marangoni stresses on the Microflow in an Evaporating Sessile Droplet," *Langmuir*, vol. 21, pp. 3972-3980, 2005.
[61] E. Widjaja, N. Liu, M. Li, R. T. Collins, O. A. Basaran, and M. T. Harris, "Dynamics of sessile drop evaporation: A comparison of the spine and the elliptic mesh generation methods," *Computers and Chemical Engineering*, vol. 31, pp. 219-232, 2007.
[62] E. Widjaja and M. T. Harris, "Particle deposition study during sessile drop evaporation," *AIChe Journal*, vol. 54, pp. 2250-2260, 2008.
[63] N. Shahidzadeh-Bonn, S. Rafai, A. Azouni, and D. Bonn, "Evaporating droplets," *Journal of Fluid Mechanics*, vol. 549, pp. 307-313, 2006.
[64] C. Poulard, G. Guena, A. M. Cazabat, A. Boudaoud, and M. B. Amar, "Rescaling the Dynamics of Evaporating Drops," *Langmuir*, vol. 21, pp. 8226-8233, 2005.
[65] E. F. Crafton and W. Z. Black, "Heat Transfer and Evaporation Rates of Small Liquid Droplets on Heated Horizontal Surfaces," *International Journal of Heat and Mass Transfer*, vol. 47, pp. 1187-1200, 2004.
[66] "www.mellesgriot.com/products/optics/mp_3_2.htm."
[67] J. Sun, "Laser-Based Thermal Pulse Measurement of Liquid Thermal Conductivity and Thermal Diffusivity," *Department of Mechanical Engineering, State University of New York at Stony Brook*, vol. Masters thesis, pp. 77, 1999.
[68] "http://www.scioncorp.com/pages/scion_image_windows.htm."
[69] R. Bhardwaj, X. Fang, and D. Attinger, "Pattern formation during the evaporation of a colloidal nanoliter drop: a numerical and experimental study," *New Journal of Physics*, vol. 11, pp. 075020, 2009.
[70] D. Attinger and D. Poulikakos, "Melting and Resolidification of a Substrate caused by Molten Microdroplet Impact," *Journal of Heat Transfer*, vol. 123, pp. 1110-1122, 2001.
[71] R. Bhardwaj, J. P. Longtin, and D. Attinger, "A numerical investigation on the influence of liquid properties and interfacial heat transfer during microdroplet deposition onto a glass substrate," *International Journal of Heat and Mass Transfer*, vol. 50, pp. 2912-2923, 2007.
[72] R. Bhardwaj and D. Attinger, "Non-isothermal wetting during impact of millimeter size water drop on a flat substrate: numerical investigation and comparison with high speed visualization experiments," *International Journal of Heat and Fluid Flow*, vol. 29, pp. 1422-1435, 2008.
[73] T. D. Blake and J. DeConinck, "The influence of solid-liquid interactions on dynamic wetting," *Advances in Colloid and Interface Science*, vol. 66, pp. 21, 2002.
[74] D. R. Lide, *CRC Handbook of Chemistry and Physics on CD-ROM*, 81st ed: Chapman and Hall/CRC, 2001.
[75] N. B. Vargaftik, *Handbook of physical properties of liquids and gases: pure substances and mixtures (p 644)*. Washington DC: Hemisphere publishing, 1975.
[76] M. E. Lusty and M. H. Dunn, "Refractive indices and thermo-optical properties of dye laser solvents," *Applied Physics B*, vol. 44, pp. 193-198, 1987.
[77] D. Solimini, "Loss measurement of organic material at 6328 Angstroms," *Journal of Applied Physics*, vol. 37, pp. 3314-3315, 1966.
[78] J. Stone, "Measurements of the absorption of light in low-loss liquids," *Journal of Optical Society of America*, vol. 62, pp. 327-333, 1972.
[79] S. Schiaffino and A. A. Sonin, "Molten Droplet Deposition and Solidification at Low Weber Numbers," *Physics of Fluids*, vol. 9, pp. 3172-3187, 1997.
[80] H. S. Carslaw and J. C. Jaeger, *Conduction of Heat in Solids*: Clarendon Press, Oxford, 1959.
[81] T. Loulou and D. Delaunay, "The interface temperature of two suddenly contacting bodies, one of them undergoing phase change," *International Journal of Heat and Masss Transfer*, vol. 40, pp. 1713-1716, 1997.
[82] Y. Popov, "Singularities, Universality and Scaling in Evaporative Deposition Patterns," PhD thesis, University of Chicago, 2003, pp. Available at: http://jfi.uchicago.edu/~tten/Coffee.drops/ (retrieved on October 2007).
30

# 8 Figure captions

Figure 1: (a) Experimental setup, with details (b) of the laser beam propagation. The laser beam is focused and reflected at the top surface of a glass prism, on which the drop impacts. The change of reflected intensity is related to the change of temperature at the interface.

Figure 2: Calibration curve for converting the laser reflectivity $R$ to the liquid temperature at the interface. $R_0$ is the reflectivity at ambient temperature and $r$ is the correlation coefficient of the linear regression. Three measurements have been performed at four different temperatures.

Figure 3a: Impact of a 3µL isopropanol droplet at ambient temperature (23°C) on a fused silica substrate heated at 68°C. Comparison of droplet shapes from experiments (left) with simulated streamlines and temperature field (right). The substrate level is shown in the left halves of the images using a white line.

Figure 3b: continuation of (a) for latter times

Figure 4: Impact of a 3 µL isopropanol droplet at ambient temperature (23°C) on a fused silica substrate heated at 68°C. Comparison of experimental and numerical results for wetted radius and maximum drop height for the same impact case as in Figure 3. Results are non-dimensionalized with respect to the initial droplet diameter. Only one error bar (±12 µm) per measurement is shown for clarity.

Figure 5: Comparison between laser temperature measurements and simulations of the temperature at the center of the drop-substrate interface for the same impact case as in Figure 3. Simulations are plotted for different values of interfacial heat transfer coefficient.

Figure 6: Influence of substrate temperature on the wetted radius and maximum height during impact of a 3 µL isopropanol droplet at ambient temperature (23°C). Impact velocity is 0.37 m s$^{-1}$. Only one average error bar (±12 µm) per measurement is shown for clarity.

Figure 7: Influence of substrate temperature on the interfacial temperature under a 3µL isopropanol droplet at ambient temperature (23°C) impacting a heated substrate. Impact velocity is 0.37 m s$^{-1}$. For each three initial substrate temperatures: $T_{sub}$ = 25, 56 and 68°C, the interfacial temperature is measured three times, indicated by the arrows and legend. For each initial temperature, three simulations are shown (dash dot dot, dashed and solid lines) corresponding to different values of interfacial heat transfer coefficient.

Figure 8: Impact and evaporation of a 74 nL isopropanol droplet at ambient temperature (23.5°C) on a fused silica substrate heated at 41°C. Comparison of experimental and numerical results for wetted radius and maximum drop height. Results are non-dimensionalized with respect to the initial droplet diameter. The drop volume is only measured when the drop has assumed a spherical cap ($t$ > 40 ms). Only one error bar per measurement is shown for clarity. The measurement of a 79 nL droplet is also shown.

Figure 9: Influence of substrate temperature on the interfacial temperature under a nL-volume isopropanol droplet at ambient temperature (23.5°C) impacting and evaporating on a heated substrate. For each four initial substrate temperatures: $T_{sub}$ = 25, 41, 49 and 57°C, the measured interfacial temperature is shown, indicated by an arrow and legend. The respective initial drop volumes in the four cases are 45, 74, 90 and 64 nL. Final evaporation times ($t_f$) measured from high-speed visualization are shown as vertical lines.

For each initial temperature, two simulations are shown (S1 and S2) respectively without and with consideration of the Marangoni convection.

Figure 10: Impact and evaporation of a 74 nL isoporopanol droplet at ambient temperature (23.5°C) on a fused silica substrate heated at 41°C. Impact velocity 0.4 m s$^{-1}$. Comparison of droplet shapes from experiments (left) with simulated streamlines and temperature field (right). The substrate level is shown in the left halves of the images using a white line.

## 9 Figures

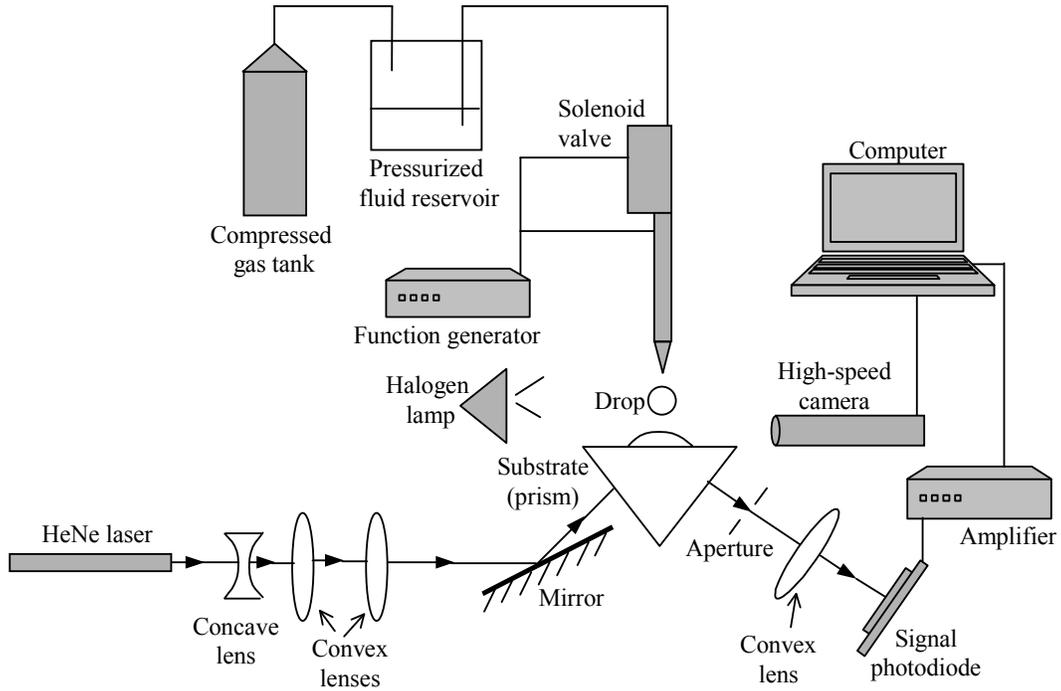

**a**

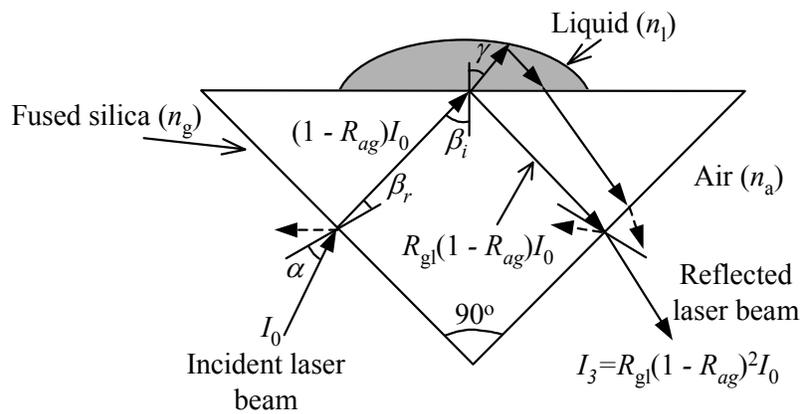

**b**

Figure 1: (a) Experimental setup, with details (b) of the laser beam propagation. The laser beam is focused and reflected at the top surface of a glass prism, on which the drop impacts. The change of reflected intensity is related to the change of temperature at the interface.

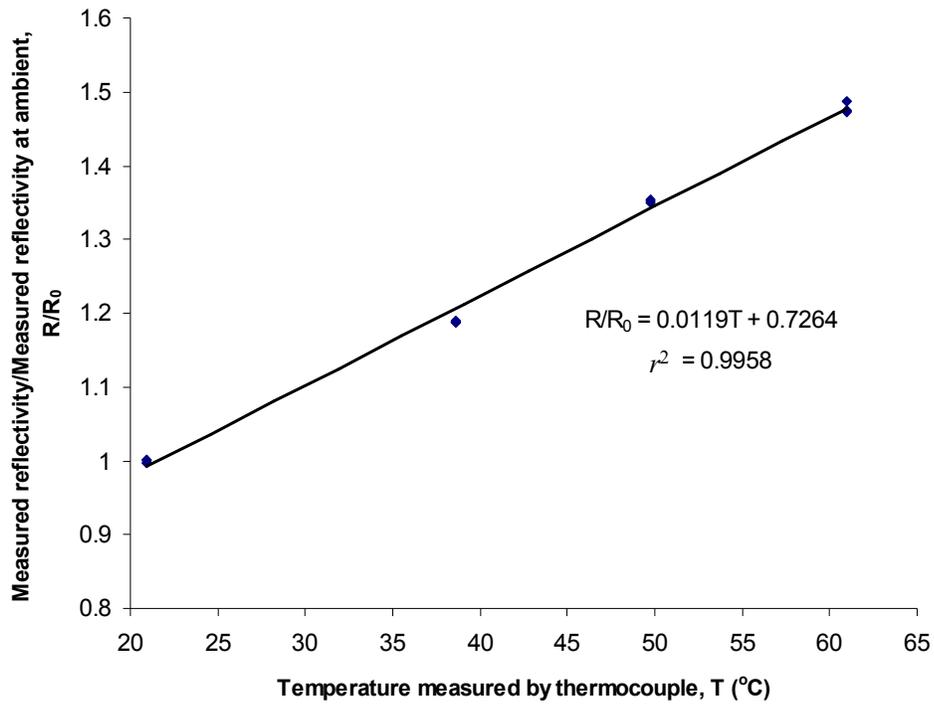

Figure 2: Calibration curve for converting the laser reflectivity $R$ to the liquid temperature at the interface. $R_0$ is the reflectivity at ambient temperature and $r$ is the correlation coefficient of the linear regression. Three measurements have been performed at four different temperatures.

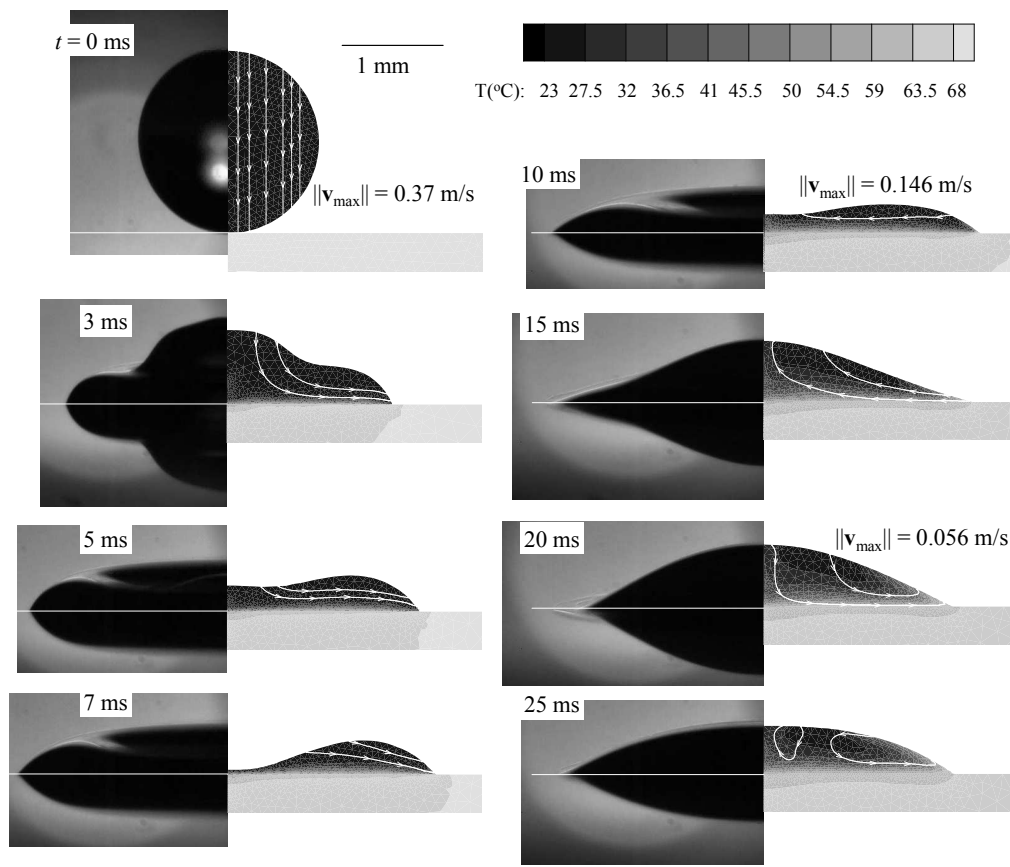

Figure 3a: Impact of a 3µL isopropanol droplet at ambient temperature (23°C) on a fused silica substrate heated at 68°C. Comparison of droplet shapes from experiments (left) with simulated streamlines and temperature field (right). The substrate level is shown in the left halves of the images using a white line.

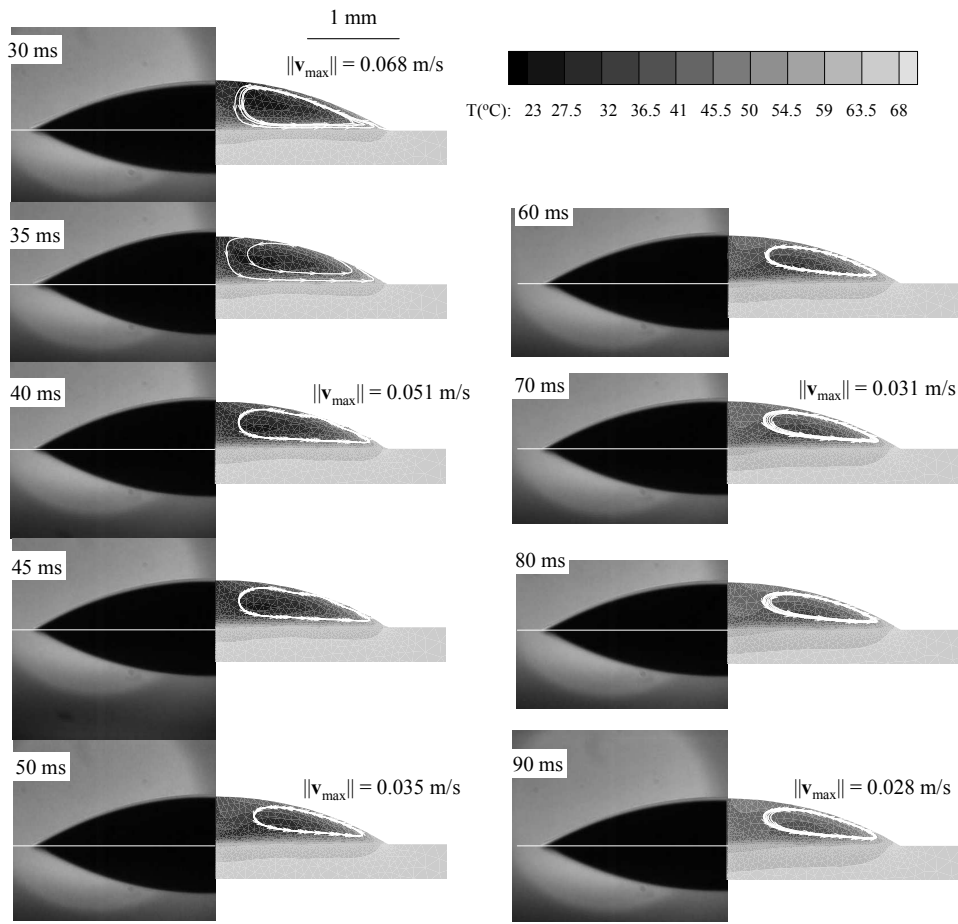

Figure 3b: continuation of (a) for latter times

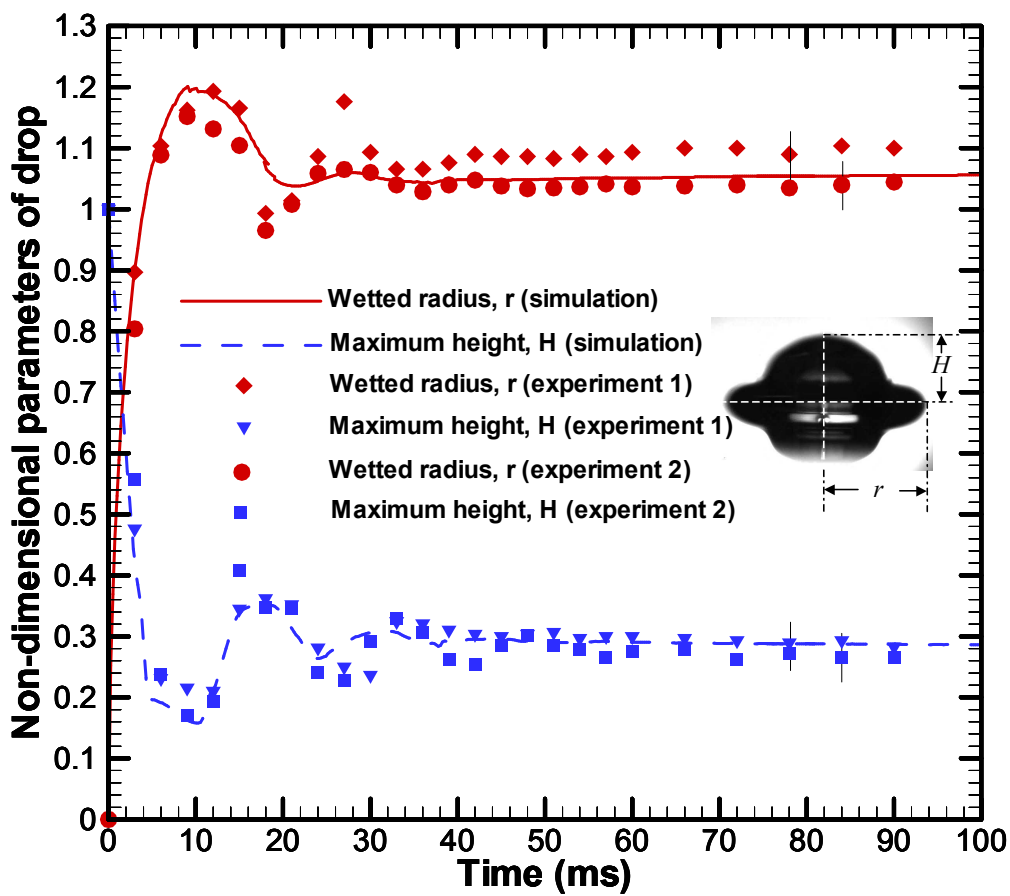

Figure 4: Impact of a 3 µL isopropanol droplet at ambient temperature (23°C) on a fused silica substrate heated at 68°C. Comparison of experimental and numerical results for wetted radius and maximum drop height for the same impact case as in Figure 3. Results are non-dimensionalized with respect to the initial droplet diameter. Only one error bar (±12 µm) per measurement is shown for clarity.

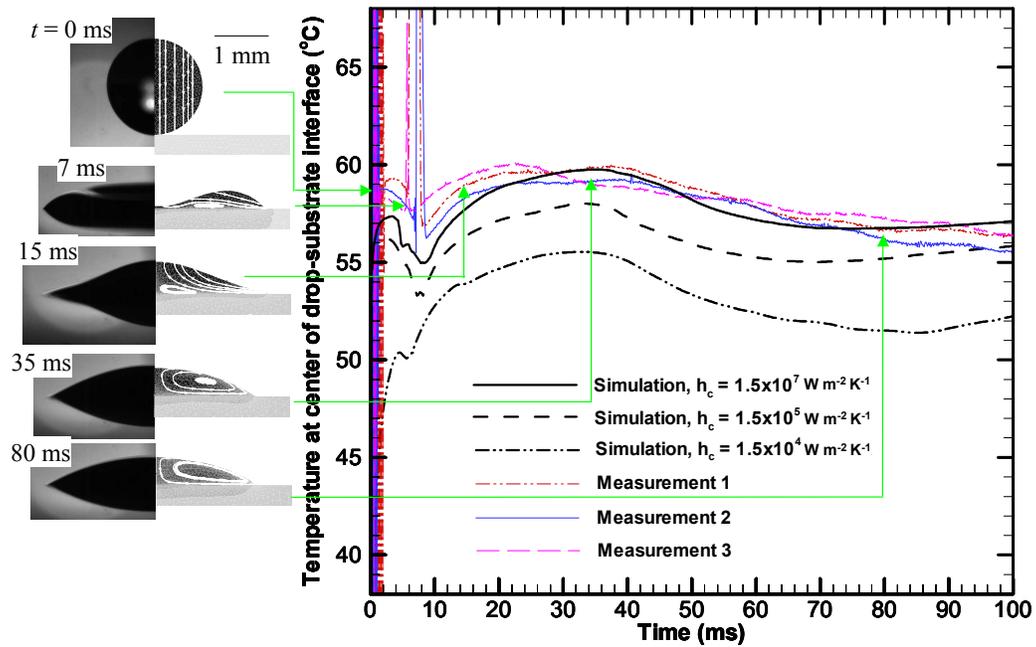

Figure 5: Comparison between laser temperature measurements and simulations of the temperature at the center of the drop-substrate interface for the same impact case as in Figure 3. Simulations are plotted for different values of interfacial heat transfer coefficient.

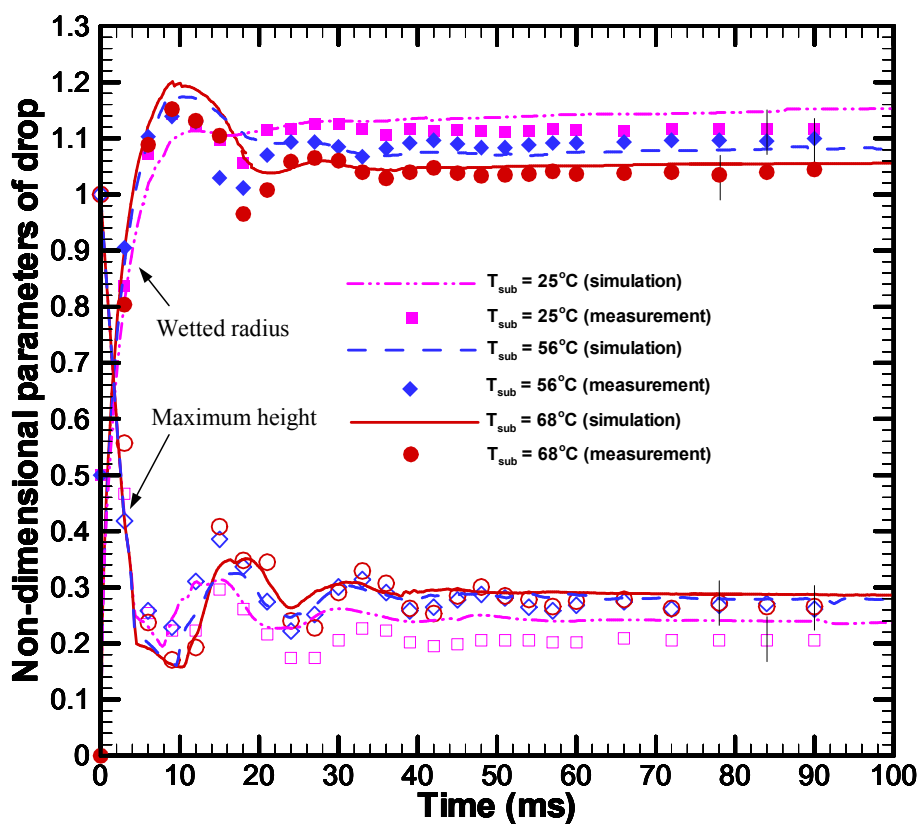

Figure 6: Influence of substrate temperature on the wetted radius and maximum height during impact of a 3 μL isopropanol droplet at ambient temperature (23°C). Impact velocity is 0.37 m s$^{-1}$. Only one average error bar (±12 μm) per measurement is shown for clarity.

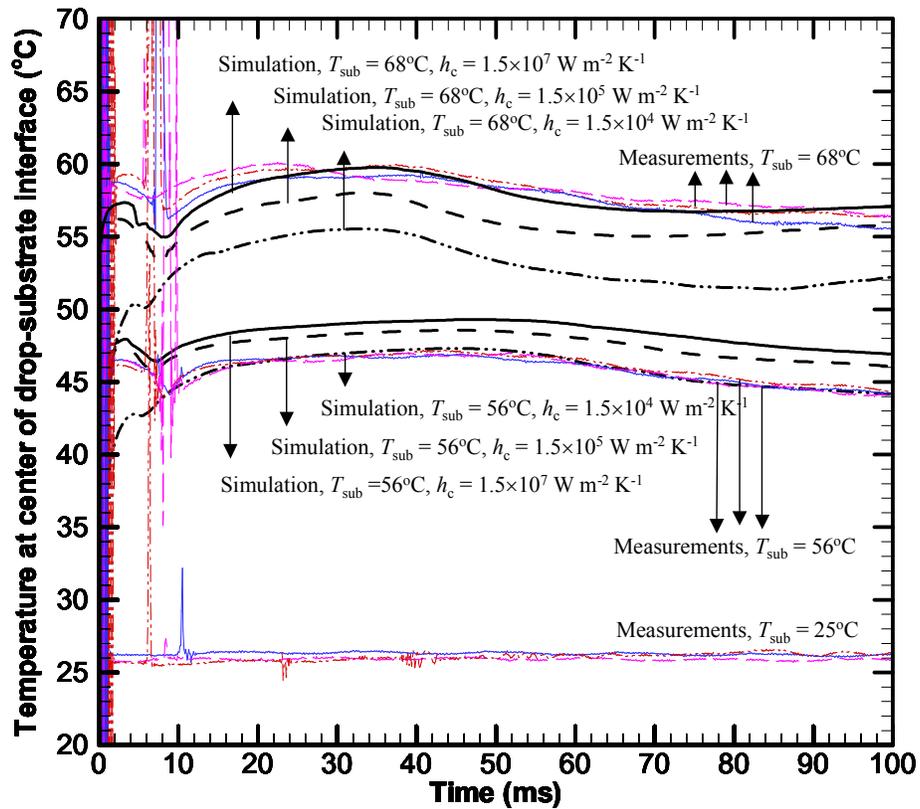

Figure 7: Influence of substrate temperature on the interfacial temperature under a 3μL isopropanol droplet at ambient temperature (23°C) impacting a heated substrate. Impact velocity is 0.37 m s$^{-1}$. For each three initial substrate temperatures: $T_{sub}$ = 25, 56 and 68°C, the interfacial temperature is measured three times, indicated by the arrows and legend. For each initial temperature, three simulations are shown (dash dot dot, dashed and solid lines) corresponding to different values of interfacial heat transfer coefficient.

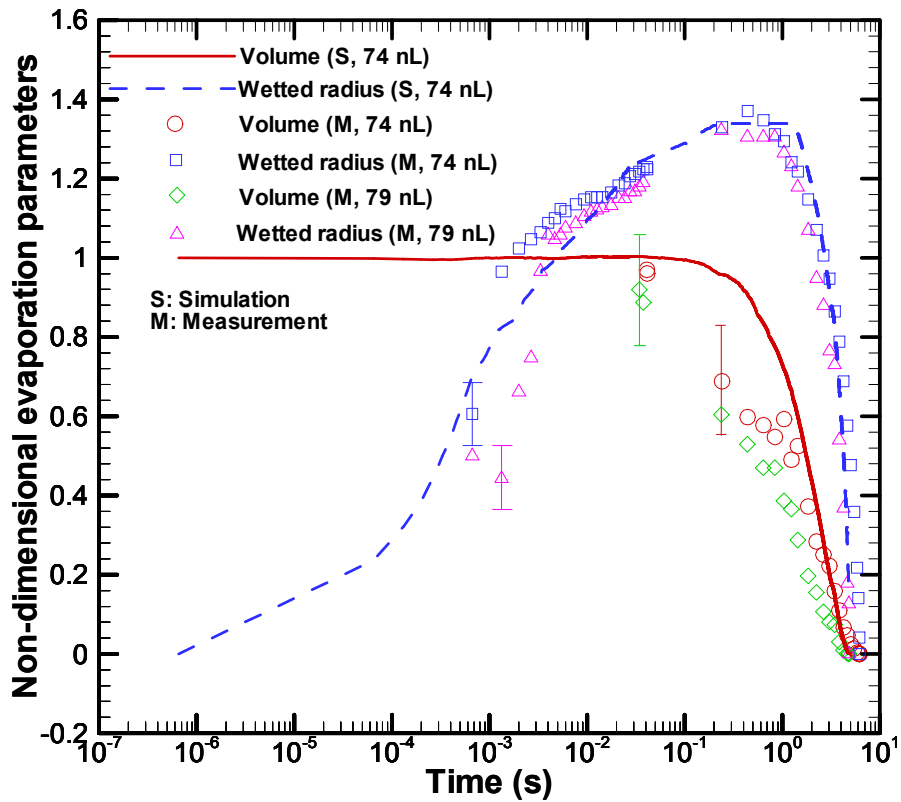

Figure 8: Impact and evaporation of a 74 nL isopropanol droplet at ambient temperature (23.5°C) on a fused silica substrate heated at 41°C. Comparison of experimental and numerical results for wetted radius and maximum drop height. Results are non-dimensionalized with respect to the initial droplet diameter. The drop volume is only measured when the drop has assumed a spherical cap ($t > 40$ ms). Only one error bar per measurement is shown for clarity. The measurement of a 79 nL droplet is also shown.

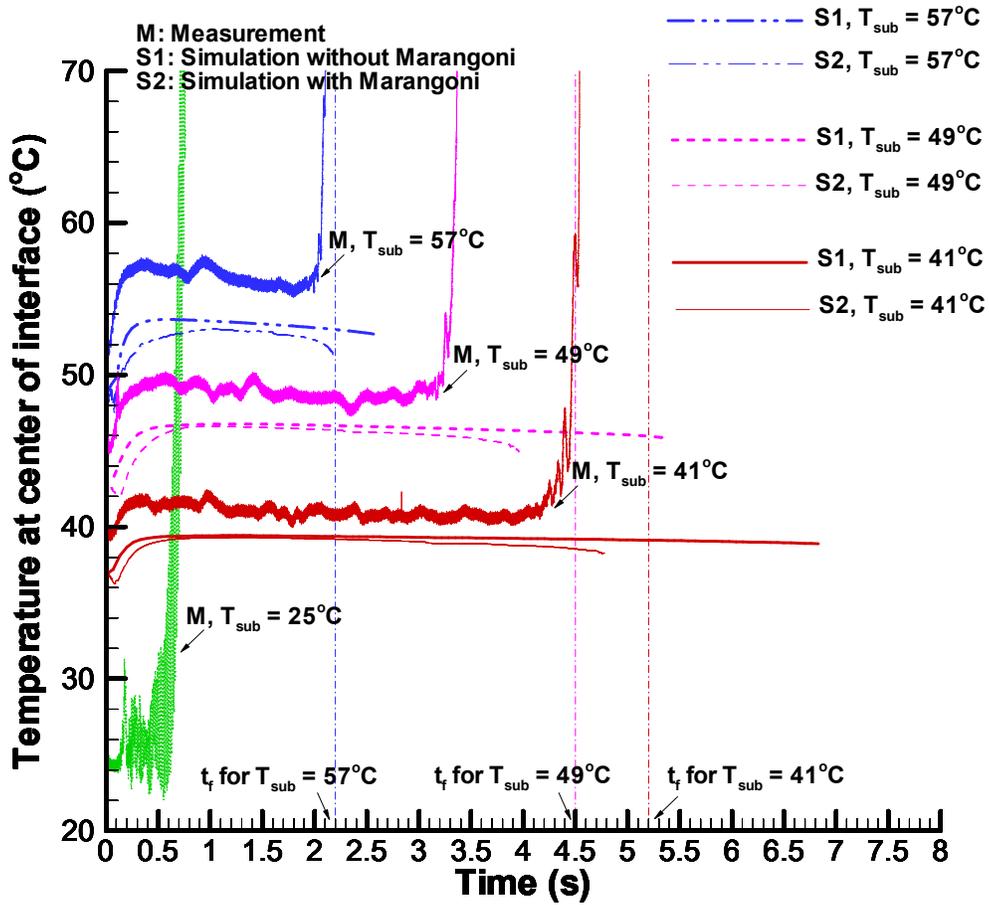

Figure 9: Influence of substrate temperature on the interfacial temperature under a nL-volume isopropanol droplet at ambient temperature (23.5°C) impacting and evaporating on a heated substrate. For each four initial substrate temperatures: $T_{sub}$ = 25, 41, 49 and 57°C, the measured interfacial temperature is shown, indicated by an arrow and legend. The respective initial drop volumes in the four cases are 45, 74, 90 and 64 nL. Final evaporation times ($t_f$) measured from high-speed visualization are shown as vertical lines. For each initial temperature, two simulations are shown (S1 and S2) respectively without and with consideration of the Marangoni convection.

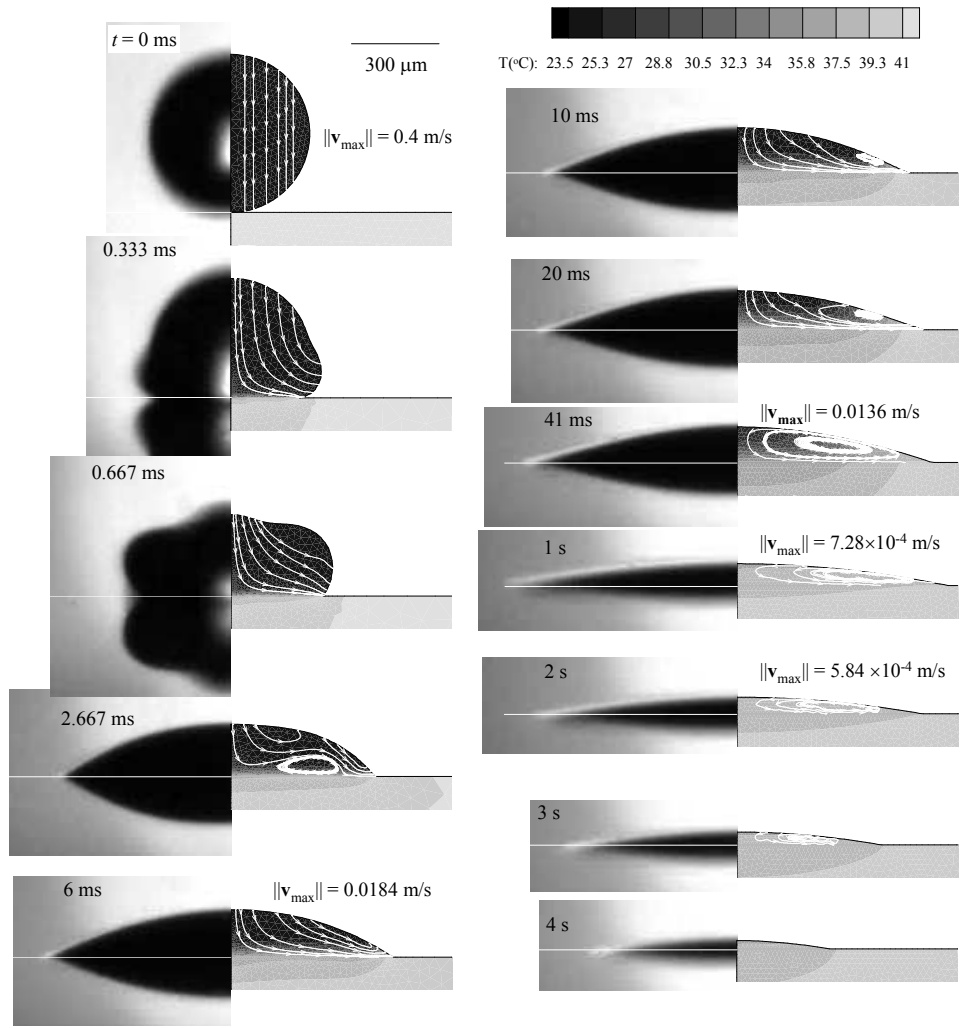

Figure 10: Impact and evaporation of a 74 nL isoporopanol droplet at ambient temperature (23.5°C) on a fused silica substrate heated at 41°C. Impact velocity 0.4 m s$^{-1}$. Comparison of droplet shapes from experiments (left) with simulated streamlines and temperature field (right). The substrate level is shown in the left halves of the images using a white line.